\documentclass[12pt]{iopart}
\usepackage{amsfonts}
\usepackage{amsbsy}

\makeatletter 
  \@addtoreset{equation}{section} 
  
\makeatother

\begin{document}

\title{B\"acklund transformations between the AKNS and DNLS hierarchies.}

\author{V.E. Vekslerchik}

\address{Institute for Radiophysics and Electronics, Kharkov, Ukraine}

\ead{vekslerchik@yahoo.com}

\begin{abstract}
Starting from the functional representation of the Ablowitz-Kaup-Newell-Segur 
(AKNS) and derivative nonlinear Schr\"odinger (DNLS) hierarchies and using 
the chains of the Miura-like transformations we derive a set of 
B\"acklund transformations that link solutions of these systems. 
It is shown that the extended AKNS and DNLS hierarchies possess common set of 
tau-functions and their connection with the Ablowitz-Ladik hierarchy is 
established. These results are another manifestation of the already known 
fact that the AKNS and DNLS hierarchies are closely related and can be viewed 
as particular cases of a more general system. 
\end{abstract}


\submitto{\JPA}

\maketitle


\newcommand{\mybacklund}[1]{\mathfrak{#1}}
\newcommand{\mysup}[1]{^{\scriptscriptstyle #1}} 
\newcommand{\mygr}[2]{\mathfrak{#1}^{\scriptscriptstyle #2}} 
\newcommand{\mytgr}[2]{\tilde\mathfrak{#1}^{\scriptscriptstyle #2}} 
\newcommand{\mygamma}[1]{\Gamma^{\scriptscriptstyle #1}} 

\section{Introduction.}

This paper is devoted to the B\"acklund transformations (BTs) between the 
Ablowitz-Kaup-Newell-Segur (AKNS) hierarchy and the derivative nonlinear 
Schr\"odinger (DNLS) hierarchy.
The relationships between these integrable systems have been studied by 
different authors (see, \textit{e.g.}, book \cite{N85} and references therein).
Moreover, the nonlinear Schr\"odinger (NLS) equation, which is the simplest 
equation of the AKNS hierarchy, and DNLS equation can be viewed as particular 
cases of a more general system. So, the authors of \cite{WKI1979,KSK80,IKWS80} 
have shown that the  generalized (or mixed, or hybrid) NLS equation, 
which has both usual cubic nonlinear term  (as in the NLS equation) and a 
derivative cubic term (as in the DNLS equation), is also an integrable system 
and elaborated the correspondent versions of the inverse scattering transform.
Later Flaschka, Newell and Ratiu demonstrated in \cite{FNR83} that both 
hierarchies (AKNS and DNLS) are parts of the greater hierarchy connected with 
$\widetilde{sl}(2,C)$. Also we would like to mention the results obtained by 
Dimakis and M\"uller-Hoissen \cite{DM06a} who recovered the DNLS hierarchy as 
AKNS-type one using the deformations of associative products.
However, usually these equations/hierarchies are treated as different and a 
natural question is to study the transformations between solutions of these 
closely related systems. 
This problem has been solved for the NLS and DNLS equations by Ishimori 
\cite{I82}, Wadati and Sogo \cite{WS83} (see also \cite{HW94} for the 
multicomponent case), who constructed the gauge transformations between the 
matrices which form the zero-curvature representation 
of the corresponding equations. 

In this work the results of \cite{I82,WS83} are extended to the level of 
hierarchies. In doing this we will not use the approach based on inverse 
scattering technique and will not develop the gauge transformations between 
the scattering problems of the hierarchies. Instead all the relationships 
are formulated directly in terms of solutions starting from the functional 
representation of the AKNS and DNLS hierarchies (instead of the 
zero-curvature one).

After presenting in section \ref{sec-hies} some facts related to the 
functional representation of the AKNS and DNLS hierarchies and their 
Miura-like transformations, we introduce and prove the B\"acklund relations 
(sections \ref{sec-bt}, \ref{sec-one-point}) which then are simplified in 
section \ref{sec-sol}. Finally, we rewrite in section \ref{sec-bilinear}  
the obtained BTs in the bilinear form and demonstrate their relations with the 
Ablowitz-Ladik hierarchy (ALH).

\section{The AKNS and DNLSE hierarchies. \label{sec-hies}}

The key feature of the presented work is the so-called functional representation 
of the hierarchies when an infinite number of differential equations is replaced 
by a few functional equations generated by the Miwa's shifts applied to the 
functions of an infinite number of arguments:
\begin{equation}
    \mathbb{E}_{\xi} Q(\mathrm{t}) = 
    Q\left( \mathrm{t} + i [\xi] \right) 
\end{equation}
where
\begin{equation}
    Q(\mathrm{t}) = 
    Q\left( t_{1}, t_{2}, ...\right) = 
    Q\left( t_{k} \right) 
\end{equation}
and 
\begin{equation}
    Q\left( \mathrm{t} + i [\xi] \right) = 
    Q\left( t_{1} + i \xi, t_{2} + i \xi^{2}/2, ... \right) = 
    Q\left( t_{k} + i \xi^{k} / k \right) 
\end{equation}
(see, \textit{e.g.}, \cite{V02,DM06b}). 
In what follows we use the functional representations of the AKNS and DNLS 
hierarchies that have been derived in \cite{V02,PV02}.

\subsection{AKNS hierarchy.}

The AKNS hierarchy was introduced in 1974 in \cite{AKNS74}. In that work the 
authors generalized the inverse scattering approach of Zakharov and Shabat that 
was developed in \cite{ZS71,ZS73} for solution of the NLS equation. This 
hierarchy consists of an infinite family of integrable equations which, in the 
framework of the inverse scattering method, can be associated with the 
Zakharov-Shabat eigenvalue problem used in \cite{ZS71,ZS73}.

The AKNS hierarchy can be written as the system of two-parametric equations 
\begin{equation}
  \left\{
  \begin{array}{lcl} 
    \mygr{A}{Q}(\xi,\eta) & = & 0 
  \\
    \mygr{A}{R}(\xi,\eta) & = & 0 
  \end{array}
  \right.
\label{hie-akns}
\end{equation}
where the generating relations $\mygr{A}{Q}(\xi,\eta)$  and 
$\mygr{A}{R}(\xi,\eta)$ are defined by 
\begin{equation}
  \begin{array}{rcl} 
    \mygr{A}{Q}(\xi,\eta) & = & 
    ( \xi - \eta ) \hat{\bar{Q}} 
    - 
    \left( 1 + \xi\eta \, \hat{\bar{Q}} R \right) 
    \left( \xi\hat{Q} - \eta\bar{Q} \right) 
  \\[2mm]
    \mygr{A}{R}(\xi,\eta) & = & 
    ( \xi - \eta ) R 
    - 
    \left( 1 + \xi\eta \, \hat{\bar{Q}} R \right) 
    \left( \xi\bar{R} - \eta\hat{R} \right) 
  \end{array}
\label{hie-akns-def}
\end{equation}
with the shortcuts that will be repeatedly used throughout the paper, 
\begin{equation}
    \hat{Q} = \mathbb{E}_{\xi} Q, 
    \qquad 
    \bar{Q} = \mathbb{E}_{\eta} Q.
\end{equation}
Equations (\ref{hie-akns}) can be simplified by different limiting procedures. 
For example, sending $\eta$ to zero one comes to a system that was used in 
\cite{PV02}:
\begin{equation}
  \left\{ 
  \begin{array}{rcl} 
  Q - \hat{Q} + i\xi \partial_{1}\hat{Q} - \xi^{2} \hat{Q}^{2} R 
  & = & 0
  \\[2mm]
  \hat{R} - R - i\xi \partial_{1} R - \xi^{2} \hat{Q} R^{2} 
  & = & 0
  \end{array}
  \right.
\label{hie-akns-B}
\end{equation}
where 
$\partial_{j} = \partial/\partial t_{j}$. 
Expanding these functional equations in the power series in $\xi$ one obtains 
an infinite set of differential ones, the first of which are the NLS equation, 
\begin{equation}
  \left\{ 
  \begin{array}{rcl} 
  i\partial_{2} Q + \partial_{11} Q + 2 Q^{2}R 
  & = & 0
  \\ 
  -i\partial_{2} R + \partial_{11} R + 2 QR^{2} 
  & = & 0
  \end{array}
  \right.
\end{equation}
and the complex mKdV equation,
\begin{equation}
  \left\{ 
  \begin{array}{rcl} 
  \partial_{3} Q + \partial_{111} Q 
  + 6 QR \partial_{1} Q 
  & = & 0
  \\ 
  \partial_{3} R + \partial_{111} R 
  + 6 QR \partial_{1} R 
  & = & 0
  \end{array}
  \right.
\end{equation}
(here $\partial_{jk}$ stand for $\partial^{2} / \partial t_{j}\partial t_{k}$, \textit{etc})
that are the first equations of the AKNS hierarchy.
However, in what follows it is more convenient to work with the `algebraic' 
equations (\ref{hie-akns}) than with the `differential' ones (\ref{hie-akns-B}).

Another key ingredient of this work is to consider, instead of a single 
solution of the hierarchy, $(Q,R)$, an infinite set of solutions 
$(Q_{n},R_{n})$, $n \in (-\infty, \infty)$ related by the 
Miura-like transformations:
\begin{equation}
  \left\{
  \begin{array}{rcl} 
  \mygr{M}{Q}_{n}(\xi) & = & 0 
  \\ 
  \mygr{M}{R}_{n}(\xi) & = & 0 
  \end{array}
  \right.
  \label{miura-akns}
\end{equation}
where
\begin{equation}
  \begin{array}{rcl} 
  \mygr{M}{Q}_{n}(\xi) & = & 
  \hat{Q}_{n} 
  - \mygamma{QR}_{n}(\xi) \left( Q_{n} + \xi \hat{Q}_{n+1} \right) 
  \\[2mm]
  \mygr{M}{R}_{n}(\xi) & = & 
  R_{n+1} 
  - \mygamma{QR}_{n}(\xi) \left( \hat{R}_{n+1} + \xi R_{n} \right) 
  \end{array}
\end{equation}
with 
\begin{equation}
  \mygamma{QR}_{n}(\xi) = 
  1 - \xi \hat{Q}_{n} R_{n+1}. 
\end{equation}
It can be shown that if a pair $(Q_{n_{0}},R_{n_{0}})$ solves equations 
(\ref{hie-akns}), then, by virtue of (\ref{miura-akns}), so do the pairs 
$(Q_{n_{0} \pm 1},R_{n_{0} \pm 1})$ and hence the pairs 
$(Q_{n},R_{n})$ for all values of $n$ (see \ref{app-lr-akns}).

\subsection{DNLS hierarchy.}

There are different forms to write the DNLS equation \cite{KN78,CLL79,GI83}. 
In this work, we use the one that was presented by Chen, Lee and Liu in 
\cite{CLL79}. 

By analogy with the AKNS case, discussed in the previous section, the 
DNLS hierarchy can be written as 
\begin{equation}
  \left\{
  \begin{array}{rcl} 
  \mygr{A}{U}(\xi,\eta) & = & 0 
  \\ 
  \mygr{A}{V}(\xi,\eta) & = & 0 
  \end{array}
  \right.
\label{hie-dnls}
\end{equation}
where
\begin{equation}
  \begin{array}{rcl} 
  \mygr{A}{U}(\xi,\eta) & = & 
  \left[ 
    \xi - \eta  
    + \xi\eta \left( \bar{U} - \hat{U} \right) V 
  \right] \hat{\bar{U}} 
  - \xi \hat{U} + \eta \bar{U} 
  \\[2mm]
  \mygr{A}{V}(\xi,\eta) & = & 
  \left[ 
    \xi - \eta  
    + \xi\eta \, \hat{\bar{U}} \left( \hat{V} - \bar{V} \right) 
  \right] V 
  + \eta \hat{V} - \xi \bar{V}  
  \end{array}
\label{hie-dnls-def}
\end{equation}
(see \cite{V02}). 
These functional equations in the $\eta \to 0$ limit become 
\begin{equation}
  \left\{
  \begin{array}{rcl} 
    i\xi \partial_{1} \hat{U} 
    + \left( 1 + \xi \hat{U} V \right) 
      \left( U - \hat{U} \right) 
    & = & 0 
  \\[2mm]
    -i\xi \partial_{1} V 
    + \left( 1 + \xi \hat{U} V \right) 
      \left( \hat{V} - V \right) 
    & = & 0 
  \end{array}
  \right.
\end{equation}
and lead to an infinite set of differential equations, the simplest of which is 
the DNLS equation:
\begin{equation}
  \left\{
  \begin{array}{rcl} 
    \left( i\partial_{2} + \partial_{11} \right) U 
    + 2i UV \partial_{1} U 
  & = & 0 
  \\[2mm]
    \left( -i\partial_{2} + \partial_{11} \right) V 
    - 2i UV \partial_{1} V 
  & = & 0. 
  \end{array}
  \right.
\end{equation}

The Miura-like transformations for the DNLS hierarchy are defined by 
\begin{equation}
  \left\{
  \begin{array}{lcl} 
  \mygr{M}{U}_{n}(\xi) 
  & = & 0 
  \\
  \mygr{M}{V}_{n}(\xi) 
  & = & 0 
  \end{array}
  \right. 
  \label{miura-dnls}
\end{equation}
with 
\begin{equation}
  \begin{array}{lcl} 
  \mygr{M}{U}_{n}(\xi) 
  & = & 
  \mygamma{UV}_{n+1}(\xi) \hat{U}_{n} 
  - \left( U_{n} + \xi \hat{U}_{n+1} \right) 
  \\[2mm]
  \mygr{M}{V}_{n}(\xi) 
  & = & 
  \mygamma{UV}_{n}(\xi) V_{n+1} 
  - \left( \hat{V}_{n+1} + \xi V_{n} \right) 
  \end{array}
\end{equation}
and
\begin{equation}
  \mygamma{UV}_{n}(\xi) = 1 + \xi \hat{U}_{n} V_{n}
\end{equation}
and can be shown to be compatible with (\ref{hie-dnls}) (see \ref{app-lr-dnls}).

\section{B\"acklund transformations between the AKNS and DNLS hierarchies. 
\label{sec-bt}}

The generating relations for the BTs between the AKNS and DNLS hierarchies can 
be presented as follows: 
\begin{equation}
  \left\{
  \begin{array}{rcl} 
  \mybacklund{G}_{n}(\xi) & = & 0 
  \\ 
  \mybacklund{h}_{n} & = & 0 
  \end{array}
  \right. 
\label{bt}
\end{equation}
where
\begin{equation}
  \begin{array}{rcl} 
  \mybacklund{G}_{n}(\xi) & = & 
  \mygamma{QR}_{n-1}(\xi) \mygamma{UV}_{n}(\xi) - 1
  \\[2mm]
  \mybacklund{h}_{n} & = & 
  1 - \left( 1 + Q_{n-1} R_{n+1} \right) \left( 1 - U_{n} V_{n+1} \right).
  \end{array}
\label{bt-def}
\end{equation}
The proof of the fact that (\ref{bt}) indeed relate solutions of 
(\ref{hie-akns}) with ones of (\ref{hie-dnls}) is based on expressing the 
quantities $\mygr{A}{Q,R}_{n}$ and $\mygr{A}{U,V}_{n}$, 
that are given by (\ref{hie-akns-def}) and (\ref{hie-dnls-def}) with 
$Q$, $R$, and $U$, $V$ being replaced by 
$Q_{n}$, $R_{n}$, and $U_{n}$, $V_{n}$, as linear combinations of 
$\mybacklund{G}_{n}$ and $\mybacklund{h}_{n}$ 
(and their $\mathbb{E}$-shifted values). 
These calculations are, on the one hand, rather simple but, on the other hand, 
are rather cumbersome and fatiguing. Here we present only the main steps 
verifying some of statements in \ref{app-grbt} and leaving the simplest ones 
without a proof.

Consider the quantities
\begin{eqnarray}
    \mygr{F}{QU}_{n} & = & 
    \hat{U}_{n+1} \left( \hat{V}_{n+1} - V_{n+1} \right) 
    + \xi \hat{Q}_{n} R_{n} 
    \\ 
    \mygr{F}{RV}_{n} & = & 
    \left( U_{n} - \hat{U}_{n} \right) V_{n} 
    + \xi \hat{Q}_{n} R_{n}. 
\end{eqnarray}
In the case when (\ref{bt}) hold 
they are related to $\mygr{A}{Q,R}$ and $\mygr{A}{U,V}$ by  
\begin{eqnarray}
  \frac{ \mygr{A}{U}_{n+1} }{ \hat{\bar{U}}_{n+1} } 
  - 
  \frac{ \mygr{A}{Q}_{n} }{ \hat{\bar{Q}}_{n} } 
  & = & 
  \xi\eta\left[ 
    \mygr{F}{QU}_{n}(\xi) 
     - \mygr{F}{QU}_{n}(\eta) 
    \right] 
\label{equ-A-QU}
  \\[2mm] 
  \frac{ \mygr{A}{V}_{n} }{ V_{n} } 
  - 
  \frac{ \mygr{A}{R}_{n} }{ R_{n} } 
  & = & 
  \xi\eta\left[ 
    \mathbb{E}_{\eta} \mygr{F}{RV}_{n}(\xi) 
    - \mathbb{E}_{\xi} \mygr{F}{RV}_{n}(\eta) 
  \right] 
\label{equ-A-RV}
\end{eqnarray}
(see equations (\ref{prop-F-c}), (\ref{prop-F-d}) of \ref{app-grbt}).
On the other hand, one can show that (\ref{bt}) imply that 
$\mygr{F}{QU}_{n}$ and $\mygr{F}{RV}_{n}$ satisfy
\begin{eqnarray}
  \mygr{F}{QU}_{n} 
  & = & 
  - \frac{ \hat{Q}_{n} }{ \mygamma{QR}_{n} } \,\mygr{M}{R}_{n} 
  = 
  - \frac{ \hat{U}_{n+1} }{ \mygamma{UV}_{n} } \,\mygr{M}{V}_{n} 
\label{equ-M-QU}
  \\ 
  \mygr{F}{RV}_{n} 
  & = & 
  - \frac{ R_{n} }{ \mygamma{QR}_{n-1} } \,\mygr{M}{Q}_{n-1} 
  = 
  - \frac{ V_{n} }{ \mygamma{UV}_{n+1} } \,\mygr{M}{U}_{n} 
\label{equ-M-RV}
\end{eqnarray}
(see equations (\ref{prop-F-a})--(\ref{prop-F-b}) of \ref{app-grbt}).
It is easy to conclude from the above formulae that
\begin{equation}
  \left\{
  \begin{array}{l} \displaystyle 
  \mybacklund{G}_{n} = \mybacklund{h}_{n} = 0 
  \\
  \mygr{A}{Q}_{n} = \mygr{A}{R}_{n} = 0 
  \end{array}
  \right\}_{n = -\infty, ... , \infty}
  \quad\Rightarrow\quad
  \left\{
  \mygr{A}{U}_{n} = \mygr{A}{V}_{n} = 0 
  \right\}_{n = -\infty, ... , \infty}
\label{lemma-QGU}
\end{equation}
and 
\begin{equation}
  \left\{
  \begin{array}{l} \displaystyle 
  \mybacklund{G}_{n} = \mybacklund{h}_{n} = 0 
  \\
  \mygr{A}{U}_{n} = \mygr{A}{V}_{n} = 0 
  \end{array}
  \right\}_{n = -\infty, ... , \infty}
  \quad\Rightarrow\quad
  \left\{
  \mygr{A}{Q}_{n} = \mygr{A}{R}_{n} = 0 
  \right\}_{n = -\infty, ... , \infty}
\label{lemma-UGQ}
\end{equation}
which means that (\ref{bt}), (\ref{bt-def}) are indeed transformations that link 
solutions of the AKNS and DNLS hierarchies.

Let us prove, for example, the first of these statements. 
The conditions from the left-hand side of (\ref{lemma-QGU}) 
together with the definitions of $Q_{n \pm 1}$ and $R_{n \pm 1}$, 
$\mygr{M}{Q}_{n} = \mygr{M}{R}_{n} = 0$, 
and identities (\ref{equ-M-QU}), (\ref{equ-M-RV}) imply that 
$\mygr{F}{QU}_{n} = \mygr{F}{RV}_{n} = 0$ for all $n$.
Then, it follows from (\ref{equ-A-QU}), (\ref{equ-A-RV}) that  
$\mygr{A}{U}_{n} = \mygr{A}{V}_{n} = 0$  ($n = -\infty, ... , \infty)$
which proves (\ref{lemma-QGU}).
The second statement, (\ref{lemma-UGQ}), can be demonstrated in a similar way.

\section{Solution of B\"acklund equations. \label{sec-sol}}

The aim of this section is to look at BTs (\ref{bt}), (\ref{bt-def}) from the 
practical viewpoint and to express solutions of the AKNS hierarchy in terms of 
solutions of the DNLS hierarchy and vice versa.
To this end consider the quantities 
\begin{eqnarray}
  \mygr{K}{QR}_{n}(\xi) 
  & = & 
  \frac{ 1 + Q_{n-1}R_{n+1} }{ \mygamma{QR}_{n} } 
  - 
  \frac{ 1 + \hat{Q}_{n-1}\hat{R}_{n+1} }{ \mygamma{QR}_{n-1} } 
\label{backlund-K-QR}
  \\[2mm] 
  \mygr{K}{UV}_{n}(\xi) 
  & = & \displaystyle  
  \frac{ 1 - U_{n}V_{n+1} }{ \mygamma{UV}_{n+1} } 
  - 
  \frac{ 1 - \hat{U}_{n}\hat{V}_{n+1} }{ \mygamma{UV}_{n} } 
\label{backlund-K-UV}
\end{eqnarray}
together with 
\begin{eqnarray}
  \mygr{L}{Q}_{n} & = & 
  \frac{ Q_{n} }{ Q_{n-1} } 
  - \frac{ U_{n+1} }{ U_{n} } \left( 1 - U_{n}V_{n+1} \right) 
\label{backlund-L-Q}
  \\[4mm] 
  \mygr{L}{R}_{n} & = & 
  \frac{ R_{n} }{ R_{n+1} } 
  - \frac{ V_{n} }{ V_{n+1} } \left( 1 - U_{n}V_{n+1} \right) 
\label{backlund-L-R}
\end{eqnarray}
and
\begin{eqnarray}
  \mygr{L}{U}_{n} & = & 
  \frac{ U_{n+1} }{ U_{n} } 
  - \frac{ Q_{n} }{ Q_{n-1} } \left( 1 + Q_{n-1}R_{n+1} \right) 
\label{backlund-L-U}
  \\[4mm] 
  \mygr{L}{V}_{n} & = & 
  \frac{ V_{n} }{ V_{n+1} } 
  - \frac{ R_{n} }{ R_{n+1} } \left( 1 + Q_{n-1}R_{n+1} \right). 
\label{backlund-L-V}
\end{eqnarray}
As is shown in \ref{app-grbt}, these functions vanish together with B\"acklund 
generating relations:
\begin{equation}
  \mybacklund{G}_{n} = \mybacklund{h}_{n} = 0 
  \quad\Rightarrow\quad
  \mygr{K}{QR}_{n} = \mygr{K}{UV}_{n} = 0, 
  \quad
  \mygr{L}{X}_{n} = 0 
  \quad ({\scriptstyle X} = 
  {\scriptstyle Q}, {\scriptstyle R}, {\scriptstyle U}, {\scriptstyle V} ).
\end{equation}
Using this fact and introducing the new functions, $F\mysup{QR}_{n}$ and 
$F\mysup{UV}_{n}$, related to $Q_{n}$, $R_{n}$ and $U_{n}$, $V_{n}$ non-locally,
\begin{equation}
  1 + Q_{n-1} R_{n+1} = \frac{ F\mysup{QR}_{n+1} }{ F\mysup{QR}_{n} }
\label{F-QR-def}
\end{equation}
and
\begin{equation}
  1 - U_{n} V_{n+1} = \frac{ F\mysup{UV}_{n} }{ F\mysup{UV}_{n+1} },
\label{F-UV-def}
\end{equation}
one can perform easy calculations leading to the solution of our problem.
Noting that 
\begin{equation}
  \mybacklund{h}_{n} = 
  \frac{ F\mysup{QR}_{n+1} }{ F\mysup{UV}_{n+1} } \,
  \frac{ F\mysup{UV}_{n} }{ F\mysup{QR}_{n} } 
  - 1 
\end{equation}
one can conclude that equation $\mybacklund{h}_{n} = 0$ implies  
\begin{equation}
  \frac{ F\mysup{UV}_{n} }{ F\mysup{QR}_{n} } = \alpha
\end{equation}
where $\alpha$ does not depend on $n$. 
Rewriting $\mygr{K}{QR}_{n}$ and $\mygr{K}{UV}_{n}$ as 
\begin{eqnarray}
  \mybacklund{K}\mysup{QR}_{n} 
  & = & 
  \frac{ \hat{F}\mysup{QR}_{n+1} }{ F\mysup{QR}_{n} } 
  \left(
    \frac{ F\mysup{QR}_{n+1} }{ \hat{F}\mysup{QR}_{n+1} \mygamma{QR}_{n} } 
    - 
    \frac{ F\mysup{QR}_{n} }{ \hat{F}\mysup{QR}_{n} \mygamma{QR}_{n-1} } 
  \right)
  \\[2mm] 
  \mybacklund{K}\mysup{UV}_{n} 
  & = & 
  \frac{ F\mysup{UV}_{n} }{ \hat{F}\mysup{UV}_{n+1} } 
  \left(
    \frac{ \hat{F}\mysup{UV}_{n+1} }{ F\mysup{UV}_{n+1} \mygamma{UV}_{n+1} } 
    - 
    \frac{ \hat{F}\mysup{UV}_{n} }{ F\mysup{UV}_{n} \mygamma{UV}_{n} } 
  \right)
\end{eqnarray}
one can easily see that
\begin{equation}
  \mygamma{QR}_{n}(\xi) = 
  \frac{ 1 }{ \gamma\mysup{QR} } \; 
  \frac{ F\mysup{QR}_{n+1} }{ \hat{F}\mysup{QR}_{n+1} }, 
\end{equation}
and
\begin{equation}
  \mygamma{UV}_{n}(\xi) = 
  \frac{ 1 }{ \gamma\mysup{UV} } \; 
  \frac{ \hat{F}\mysup{UV}_{n} }{ F\mysup{UV}_{n} } 
\end{equation}
where, again, $\gamma\mysup{QR}$ and $\gamma\mysup{UV}$ are constants with respect to $n$.
Further, the functions $\mygr{L}{X}_{n}$ can be presented as 
\begin{eqnarray}
  \mygr{L}{Q}_{n} 
  & = & 
  \frac{ U_{n+1} }{ Q_{n-1} F\mysup{UV}_{n+1} } 
  \left(
    \frac{ Q_{n} F\mysup{UV}_{n+1} }{ U_{n+1} } 
    - 
    \frac{ Q_{n-1} F\mysup{UV}_{n} }{ U_{n} } 
  \right)
  \\[6mm] 
  \mygr{L}{R}_{n} 
  & = & 
  \frac{ V_{n} F\mysup{UV}_{n} }{ R_{n+1} } 
  \left(
    \frac{ R_{n} }{ V_{n} F\mysup{UV}_{n} } 
    - 
    \frac{ R_{n+1} }{ V_{n+1} F\mysup{UV}_{n+1} } 
  \right)
\end{eqnarray}
and
\begin{eqnarray}
  \mygr{L}{U}_{n} 
  & = & 
  \frac{ Q_{n} F\mysup{QR}_{n+1} }{ U_{n} } 
  \left(
    \frac{ U_{n+1} }{ Q_{n} F\mysup{QR}_{n+1}} 
    - 
    \frac{ U_{n} }{ Q_{n-1} F\mysup{QR}_{n}} 
  \right)
  \\[6mm] 
  \mygr{L}{V}_{n} 
  & = & \displaystyle 
  \frac{ R_{n} }{ V_{n+1} F\mysup{QR}_{n} } 
  \left(
    \frac{ V_{n} F\mysup{QR}_{n} }{ R_{n} } 
    - 
    \frac{ V_{n+1} F\mysup{QR}_{n+1} }{ R_{n+1} } 
  \right).
\end{eqnarray}
Hence, equations 
$\mygr{L}{X}_{n} = 0$ for 
${\scriptstyle X} = 
{\scriptstyle Q}, {\scriptstyle R}, {\scriptstyle U}, {\scriptstyle V}$ 
lead to 
\begin{equation}
  Q_{n} = \beta\mysup{Q} \; U_{n+1} / F\mysup{UV}_{n+1},
  \quad
  R_{n} = \beta\mysup{R} \; V_{n} F\mysup{UV}_{n}
\end{equation}
and
\begin{equation}
  U_{n} = \beta\mysup{U} \; Q_{n-1} F\mysup{QR}_{n},
  \quad
  V_{n} = \beta\mysup{V} \; R_{n} / F\mysup{QR}_{n}.
\end{equation}
At last, analysis of other consequences of equations (\ref{bt})--(\ref{bt-def}) 
impose some restrictions on the constants $\alpha\mysup{X}$, $\beta\mysup{X}$ 
and $\gamma\mysup{X}$ (which are presented without derivation):
\begin{equation}
  \beta\mysup{Q}(\mathrm{t}) \beta\mysup{R}(\mathrm{t}) = 
  \beta\mysup{U}(\mathrm{t}) \beta\mysup{V}(\mathrm{t}) = 1 
\end{equation}
\begin{equation}
  \gamma\mysup{UV}(\mathrm{t},\xi) = 
  \frac{ \mathbb{E}_{\xi} \beta\mysup{Q}(\mathrm{t}) }{ \beta\mysup{Q}(\mathrm{t}) }
  = 
  \frac{ \beta\mysup{R}(\mathrm{t}) }{ \mathbb{E}_{\xi} \beta\mysup{R}(\mathrm{t}) } 
\end{equation}
\begin{equation}
  \gamma\mysup{QR}(\mathrm{t},\xi) = 
  \frac{ \mathbb{E}_{\xi} \beta\mysup{U}(\mathrm{t}) }{ \beta\mysup{U}(\mathrm{t}) }
  = 
  \frac{ \beta\mysup{V}(\mathrm{t}) }{ \mathbb{E}_{\xi} \beta\mysup{V}(\mathrm{t}) } 
\end{equation}
and 
\begin{equation}
  \alpha(\mathrm{t}) = 
  \beta\mysup{Q}(\mathrm{t}) \beta\mysup{U}(\mathrm{t}) 
  = 
  \frac{ 1 }{ \beta\mysup{R}(\mathrm{t}) \beta\mysup{V}(\mathrm{t}) }. 
\end{equation}
Choosing of all the functions $\alpha\mysup{X}$, $\beta\mysup{X}$ 
and $\gamma\mysup{X}$ two independent ones, say, $\alpha(\mathrm{t})$ 
and $\beta(\mathrm{t})=\beta\mysup{Q}(\mathrm{t})$ one comes to the final formulae: 
\begin{eqnarray}
  Q_{n} & = & \beta \; U_{n+1}/ F\mysup{UV}_{n+1} 
\label{sol-lin-Q}
  \\
  R_{n} & = & 
  \frac{ 1 }{ \beta } \; V_{n} \, F\mysup{UV}_{n} 
  \\
  F\mysup{QR}_{n} & = & 
  \frac{ 1 }{ \alpha} \; F\mysup{UV}_{n} 
\label{sol-lin-S}
\end{eqnarray}
and 
\begin{eqnarray}
  U_{n} & = & 
  \frac{ \alpha }{ \beta } \; Q_{n-1} \, F\mysup{QR}_{n} 
  \\
  V_{n} & = & 
  \frac{ \beta }{ \alpha } \; R_{n} / F\mysup{QR}_{n} 
  \\
  F\mysup{UV}_{n} & = & \alpha \; F\mysup{QR}_{n} 
\label{sol-lin-W}
\end{eqnarray}
that complete `solution' of the B\"acklund equations (\ref{bt}), (\ref{bt-def}). 

\section{B\"acklund transformations, bilinearization and the ALH. 
\label{sec-bilinear}}

In this section, we present some bilinear equations related to the problem we are 
dealing with. However, they are  not the traditional bilinear BTs which are 
bilinear equations relating the $\tau$-functions of two hierarchies. 
Instead, it is shown that the extended AKNS and DNLS hierarchies ('extended' 
means considered for an infinite set of solutions related by the Miura-like 
transformations) admit \emph{common} set of $\tau$-functions. In terms of these 
$\tau$-functions  the action of BTs obtained above is 
reduced to simple changes of the index $n$. 

Let us start with the final formulae of the previous section. 
The structure of (\ref{sol-lin-Q})--(\ref{sol-lin-W}) suggests the following 
parametrization:
\begin{equation}
  Q_{n} = \kappa\mysup{Q} \frac{ \sigma_{n+1} }{ \tau_{n} }, 
  \quad
  R_{n} = \frac{1}{\kappa\mysup{Q}} \frac{ \rho_{n-1} }{ \tau_{n} } 
\label{tau-QR}
\end{equation}
and
\begin{equation}
  U_{n} = \kappa\mysup{U} \frac{ \sigma_{n} }{ \tau_{n} }, 
  \quad
  V_{n} = \frac{1}{\kappa\mysup{U}} \frac{ \rho_{n-1} }{ \tau_{n-1} }. 
\label{tau-UV}
\end{equation}
Rewriting definitions (\ref{F-QR-def}), (\ref{F-UV-def}) as 
\begin{eqnarray} 
  \frac{ F\mysup{QR}_{n+1} }{ F\mysup{QR}_{n} } = 
  1 + Q_{n-1}R_{n+1} & = &  
  \frac{ 1 }{ \tau_{n-1}\tau_{n+1} }
  \left( \tau_{n}^{2} + \Delta^{\tau}_{n} \right) 
  \\
  \frac{ F\mysup{UV}_{n} }{ F\mysup{UV}_{n+1} } = 
  1 - U_{n}V_{n+1} & = &  
  \frac{ 1 }{ \tau_{n}^{2} }
  \left( \tau_{n-1}\tau_{n+1} - \Delta^{\tau}_{n} \right) 
\end{eqnarray}
with 
\begin{equation}
  \Delta^{\tau}_{n} = 
  \tau_{n-1}\tau_{n+1} + \rho_{n} \sigma_{n} - \tau_{n}^{2} 
\end{equation}
one can see that they are compatible with (\ref{sol-lin-S}) and (\ref{sol-lin-W}) 
and lead to 
\begin{eqnarray} 
  F\mysup{QR}_{n}  & = &  \kappa \frac{ \tau_{n-1} }{ \tau_{n} } 
  \\
  F\mysup{UV}_{n}  & = &  \alpha\kappa \frac{ \tau_{n-1} }{ \tau_{n} } 
\end{eqnarray}
provided $\Delta^{\tau}_{n} = 0$.
Calculating $\mygamma{QR}_{n}(\xi)$ and $\mygamma{UV}_{n}(\xi)$,
\begin{eqnarray}
  \mygamma{QR}_{n} & = &  
  \frac{ \hat\kappa\mysup{Q} }{ \kappa\mysup{Q} } 
  \left[
  \frac{ \kappa\mysup{U} }{ \hat\kappa\mysup{U} } 
  \,
  \frac{ \tau_{n}\hat\tau_{n+1} }{ \tau_{n+1}\hat\tau_{n} } 
  + 
  \frac{ \Delta^{\sigma}_{n+1} }{ \tau_{n+1}\hat\tau_{n} } 
  \right]
  \\[6mm]
  \mygamma{UV}_{n} & = &  
  \frac{ \hat\kappa\mysup{U} }{ \kappa\mysup{U} } 
  \left[
  \frac{ \kappa\mysup{Q}  }{ \hat\kappa\mysup{Q} } 
  \,
  \frac{ \tau_{n}\hat\tau_{n-1} }{ \tau_{n-1}\hat\tau_{n} } 
  - 
  \frac{ \Delta^{\sigma}_{n} }{ \tau_{n-1}\hat\tau_{n} } 
  \right]
\end{eqnarray}
where
\begin{equation}
  \Delta^{\sigma}_{n}(\xi) = 
  \frac{ \kappa\mysup{Q} }{ \hat\kappa\mysup{Q} } \tau_{n}\hat\tau_{n-1} 
  - 
  \frac{ \kappa\mysup{U} }{ \hat\kappa\mysup{U} } \tau_{n-1}\hat\tau_{n} 
  - 
  \xi \rho_{n-1} \hat\sigma_{n} 
\end{equation}
and substituting the result in (\ref{bt-def}) one can obtain
\begin{equation}
  \mybacklund{G}_{n}(\xi) = 
  \frac{ \hat\kappa\mysup{Q} \hat\kappa\mysup{U} }
       { \kappa\mysup{Q} \kappa\mysup{U} } 
  \frac{ \xi \rho_{n-1}\hat\sigma_{n} }
       { \tau_{n-1} \tau_{n} \hat\tau_{n-1} \hat\tau_{n} } 
  \; \Delta^{\sigma}_{n}(\xi) 
\end{equation} 
and
\begin{equation}
  \mybacklund{h}_{n} = 
  \frac{ \rho_{n}\sigma_{n} }{ \tau_{n-1} \tau_{n}^{2} \tau_{n+1} } 
  \; \Delta^{\tau}_{n}. 
\end{equation}
Thus, the the relationship between solutions of the AKNS and DNLS hierarchies 
can be re-formulated as presentation (\ref{tau-QR}), (\ref{tau-UV}) together 
with the conditions 
\begin{equation}
  \Delta^{\tau}_{n} = \Delta^{\sigma}_{n}(\xi) = 0. 
\label{bt-dd}
\end{equation}
Rewriting the functions generating Miura-like transformations, 
$\mygr{M}{X}_{n}(\xi)$, and then 
$\mygr{A}{X}_{n}(\xi,\eta)$ one comes to a set of bilinear functional 
equations, similar to (\ref{bt-dd}). These equations, which are not presented 
here, are nothing but equations forming the functional representation of the 
ALH \cite{V02}. Thus, the simplest way to summarize the above results 
is to say that solutions of the AKNS and DNLS hierarchies can be constructed 
of the same set of $\tau$-functions (the ALH $\tau$-functions) by the rule given 
by (\ref{tau-QR}) and (\ref{tau-UV}).

\section{One-point transformations. \label{sec-one-point}}

The question discussed in this section is the `traditional' BTs that do 
not use the lattice representation (based on the Miura-like transformations) of 
hierarchies, but relate \emph{one} solution of the AKNS hierarchy, ($Q$,$R$), 
with \emph{one} solution of the DNLS hierarchy, ($U$,$V$). We will not re-derive 
these transformations from scratch, but use the results presented in the previous 
sections. The main idea can be explained as follows. Among many generating 
relations determining the BTs 
($\mybacklund{G}_{n}$, $\mybacklund{h}_{n}$, 
$\mygr{F}{X}_{n}$, $\mygr{K}{XY}_{n}$, $\mygr{L}{X}_{n}$) 
one can select a subset that links ($U_{n}$,$V_{n}$) with ($Q_{n'}$,$R_{n'}$) 
only, without invoking ($Q_{n' \pm 1}$,$R_{n' \pm 1}$).
In doing this one can find the two possibilities: 
(i) transformations that connect ($U_{n}$,$V_{n}$) with ($Q_{n}$,$R_{n}$) and 
(ii) transformations that connect ($U_{n}$,$V_{n}$) with ($Q_{n-1}$,$R_{n-1}$).

\subsection{$\left( Q_{n},R_{n} \right) \leftrightarrow 
\left( U_{n},V_{n} \right) $ transformations.}

The first kind of transformations follows from the identities
\begin{eqnarray}
  \frac{ \mygr{A}{V}_{n}(\xi,\eta) }{ V_{n} } 
  - 
  \frac{ \mygr{A}{R}_{n}(\xi,\eta) }{ R_{n} } 
  & = & 
  \xi\eta \left[ 
    \mathbb{E}_{\eta} \mygr{F}{RV}_{n}(\xi) 
    - 
    \mathbb{E}_{\xi} \mygr{F}{RV}_{n}(\eta) 
  \right] 
\label{equ-A-F}
\\&& 
  + 
  \xi \mytgr{I}{R}_{n}(\eta) 
  - 
  \eta \mytgr{I}{R}_{n}(\xi) 
\nonumber
\\[4mm]
  \mygr{D}{UV}_{n}(\xi,\eta) 
  - 
  \mygr{D}{QR}_{n}(\xi,\eta) 
  & = & 
  \left( \mathbb{E}_{\eta} - 1 \right) \mygr{F}{RV}_{n}(\xi) 
  + 
  \left( 1 - \mathbb{E}_{\xi} \right)  \mygr{F}{RV}_{n}(\eta) 
\label{equ-D-F}
\end{eqnarray}
where 
\begin{eqnarray}
  \mygr{D}{QR}_{n}(\xi,\eta) & = & 
  \hat{\bar{Q}}_{n} 
  \left( \eta\hat{R}_{n} - \xi\bar{R}_{n} \right) 
  + 
  \left( \xi\hat{Q}_{n} - \eta\bar{Q}_{n} \right) R_{n} 
\label{zero-D-QR}
\\
  & = &  
  \frac{ 1 }{ \xi - \eta } 
  \left\{ 
    \left( \eta\hat{R}_{n} - \xi\bar{R}_{n} \right) 
    \mygr{A}{Q}_{n}(\xi,\eta) 
    + 
    \left( \xi\hat{Q}_{n} - \eta\bar{Q}_{n} \right) 
    \mygr{A}{R}_{n}(\xi,\eta) 
  \right\}, 
  \nonumber 
  \\[4mm]
  \mygr{D}{UV}_{n}(\xi,\eta) & = & 
  \left( \mathbb{E}_{\xi} - 1 \right) \left(\bar{U}_{n} - U_{n} \right) V_{n} 
  + 
  \left( 1 - \mathbb{E}_{\eta} \right) \left(\hat{U}_{n} - U_{n} \right) V_{n} 
\label{zero-D-UV}
  \\[2mm]
  & = & 
  \frac{ 1 }{ \xi - \eta } 
  \left\{ 
    \left( \hat{V}_{n} - \bar{V}_{n} \right) \mygr{A}{U}_{n}(\xi,\eta) 
    + 
    \left( \hat{U}_{n} - \bar{U}_{n} \right) \mygr{A}{V}_{n}(\xi,\eta) 
  \right\} 
  \nonumber 
\end{eqnarray}
and the quantities $\mytgr{I}{R}_{n}(\xi)$ are defined by
\begin{equation}
  \mytgr{I}{R}_{n} = 
  \frac{ \hat{R}_{n} }{ R_{n} } 
  - \mygamma{UV}_{n} \frac{ \hat{V}_{n} }{ V_{n} }. 
\end{equation}
Both $\mygr{F}{RV}_{n}(\xi)$ and $\mytgr{I}{R}_{n}(\xi)$ do not invoke 
$Q_{n \pm 1}$, $R_{n \pm 1}$ and vanish when (\ref{bt}) hold. 
At the same time the above equations (\ref{equ-A-F}), (\ref{equ-D-F}) 
demonstrate that if 
$\mygr{F}{RV}_{n}(\xi) = \mytgr{I}{R}_{n}(\xi) = 0$ and
$\mygr{A}{Q}_{n} = \mygr{A}{R}_{n} = 0$, then 
$\mygr{A}{U}_{n} = \mygr{A}{V}_{n} = 0$ and vice versa 
($\mygr{F}{RV}_{n}(\xi) = \mytgr{I}{R}_{n}(\xi) = 0$ and
$\mygr{A}{U}_{n} = \mygr{A}{V}_{n} = 0$ ensure 
$\mygr{A}{Q}_{n} = \mygr{A}{R}_{n} = 0$ ) 
for any $n$. 
To summarize, rewriting $\mygr{F}{RV}_{n}(\xi)$ and $\mytgr{I}{R}_{n}(\xi)$ 
with the index $n$ being omitted one can state that transformations 
defined by 
\begin{equation}
  \left\{
  \begin{array}{l}
  \xi \left( \mathbb{E}_{\xi}Q \right) R 
  = 
  \left[ \left( \mathbb{E}_{\xi}U \right) - U \right] V 
  \\[2mm] \displaystyle
  \frac{ \mathbb{E}_{\xi} R }{ R } 
  = 
  \left[ 1 + \xi \left( \mathbb{E}_{\xi}U \right) V \right] 
  \frac{ \mathbb{E}_{\xi}V }{ V } 
  \end{array}
  \right.
\end{equation}
are the BTs between the AKNS and DNLS hierarchies.
In the $\xi \to 0$ limit one comes to the transformations between solutions 
of the NLS and DNLS equations similar to ones derived by Wadati and Sogo in 
\cite{WS83}.

\subsection{$\left( Q_{n-1},R_{n-1} \right) \leftrightarrow 
\left( U_{n},V_{n} \right) $ transformations.}

In a similar way, one can restrict himself to the functions 
$\mygr{F}{QU}_{n-1}(\xi)$ and $\mytgr{I}{Q}_{n-1}(\xi)$, defined by 
\begin{equation}
  \mytgr{I}{Q}_{n} = 
  \frac{ Q_{n} }{ \hat{Q}_{n} } 
  - \mygamma{UV}_{n+1} \frac{ U_{n+1} }{ \hat{U}_{n+1} }. 
\end{equation}
They vanish when (\ref{bt}) hold and ensure 
the linear relations between $\mygr{A}{U,V}_{n}$ and $\mygr{A}{Q,R}_{n-1}$:
\begin{eqnarray}
  \frac{ \mygr{A}{U}_{n}(\xi,\eta) }{ \hat{\bar{U}}_{n} } 
  - 
  \frac{ \mygr{A}{Q}_{n-1}(\xi,\eta) }{ \hat{\bar{Q}}_{n-1} } 
  & = & 
  \xi\eta \left[ 
    \mygr{F}{QU}_{n-1}(\xi) - \mygr{F}{QU}_{n-1}(\eta)
  \right] 
\\&& 
  + \xi \mathbb{E}_{\xi}  \mytgr{I}{Q}_{n-1}(\eta) 
  - \eta\mathbb{E}_{\eta} \mytgr{I}{Q}_{n-1}(\xi) 
\nonumber
\\[4mm]
  \mygr{D}{UV}_{n}(\xi,\eta) 
  - 
  \mygr{D}{QR}_{n-1}(\xi,\eta) 
  & = & 
  \left( \mathbb{E}_{\eta} - 1 \right) \mygr{F}{QU}_{n-1}(\xi) 
  + 
  \left( 1 - \mathbb{E}_{\xi} \right) \mygr{F}{QU}_{n-1}(\eta). 
\end{eqnarray}
Thus, rewriting the definitions of 
$\mygr{F}{QU}_{n-1}(\xi)$ and $\mytgr{I}{Q}_{n-1}(\xi)$ 
in terms of $Q=Q_{n-1}$, $R=R_{n-1}$ and $U=U_{n}$, $V=V_{n}$ as 
\begin{equation}
  \left\{
  \begin{array}{l}
    \xi \left( \mathbb{E}_{\xi}Q \right) R = 
    \left( \mathbb{E}_{\xi}U \right) 
    \left[ V - \left( \mathbb{E}_{\xi}V \right) \right] 
  \\[2mm] \displaystyle
  \frac{ \mathbb{E}_{\xi}Q }{ Q } = 
  \frac{ 1 }{ 1 + \xi \left( \mathbb{E}_{\xi}U \right) V } \, 
  \frac{ \mathbb{E}_{\xi}U }{ U } 
  \end{array}
  \right.
\end{equation}
one comes to the second BT between the DNLS and AKNS hierarchies.

\section{Conclusion.}

To conclude we would like to mention some important questions that have not been 
discussed in this paper. The first problem is related to the generalized NLS 
equation. We know the relations between solutions of the NLS and DNLS equations. 
Also we know that both equations are particular cases of the generalized NLS 
equation. So it is interesting to obtain the functional representation of the 
generalized NLS hierarchy together with the reductions to the AKNS and DNLS 
cases. In other words, this problem can be re-formulated as to find the 
`interpolation' between (\ref{tau-QR}) and (\ref{tau-UV}), the task which 
seems to be not trivial. 

Another very important question is related to the involution, or complex 
conju\-gation. Usually in physical applications both NLS and DNLS equations 
appear in the form when $R$$=$$\pm\overline{Q}$ and $V$$=$$\pm\overline{U}$ 
(here overline stands for the complex conjugation). This reduction changes 
many of the properties of the hierarchies. 
Most noticeable manifestation of this fact one can get considering the 
Miura-like transformations: apparently they destroy the involution. So one 
has to make additional efforts in order to apply results of this paper to the 
physical situations. 
An example of how that can be done can be found in \cite{V11} where 
the author was dealing with the equation closely related to the DNLS hierarchy. 
More general approach to the problem of the involution is based on the 
incorporation of the so-called `negative flows'. This question that is outside the 
scope of this paper surely deserves separate studies.

\appendix

\section{Lattice representation of the AKNS hierarchy. \label{app-lr-akns}}

In this appendix, one can find the identities related to the functional 
representation of the AKNS hierarchy and its Miura-like transformations that 
provide proof of the implication 
$\mygr{A}{Q,R}_{n} \Rightarrow \mygr{A}{Q,R}_{n \pm 1}$ as well as of some other 
statements of this paper.

Consider the following linear combinations of the functions $\mygr{M}{Q,R}_{n}$:
\begin{eqnarray}
  \mytgr{M}{Q}_{n}(\xi,\eta) & = & 
  \xi\mathbb{E}_{\xi} \, \mygr{M}{Q}_{n}(\eta) 
  - 
  \eta\mathbb{E}_{\eta} \, \mygr{M}{Q}_{n}(\xi) 
\label{mmm-QR-a}
  \\
  \mytgr{M}{R}_{n}(\xi,\eta) & = & 
  \xi \mygr{M}{R}_{n}(\eta) 
  - 
  \eta \mygr{M}{R}_{n}(\xi) 
\end{eqnarray} 
and
\begin{equation}
  \mathring\mygr{M}{QR}_{n}(\xi) = 
  \frac{ 1 }{ \mygamma{QR}_{n}(\xi) } 
  \left[ R_{n+1} \mygr{M}{Q}_{n}(\xi) 
       - \hat{Q}_{n} \mygr{M}{R}_{n}(\xi) \right]. 
\label{mmm-QR-c}
\end{equation}
By straightforward calculations, one can demonstrate the identities
\begin{equation}
  R_{n+1} \mygr{A}{Q}_{n}(\xi,\eta) 
  + 
  \hat{\bar{Q}}_{n} \mygr{A}{R}_{n+1}(\xi,\eta) 
  = 
  R_{n+1} \mytgr{M}{Q}_{n}(\xi,\eta) 
  + 
  \hat{\bar{Q}}_{n} \mytgr{M}{R}_{n}(\xi,\eta) 
\label{zero-AM-QR-a}
\end{equation}
and
\begin{equation}
  \mygr{D}{QR}_{n}(\xi,\eta) - \mygr{D}{QR}_{n+1}(\xi,\eta) 
  = 
  \left( \mathbb{E}_{\xi} - 1 \right) \mathring\mygr{M}{QR}_{n}(\eta) 
  + 
  \left( 1 - \mathbb{E}_{\eta} \right) \mathring\mygr{M}{QR}_{n}(\xi) 
\label{zero-AM-QR-b}
\end{equation}
where $\mygr{D}{QR}_{n}$ is defined by (\ref{zero-D-QR}).
One can easily see that 
\begin{equation}
  \mygr{M}{Q,R}_{n} = 0 
  \quad\Rightarrow\quad
  \mytgr{M}{Q,R}_{n} = \mathring\mygr{M}{QR}_{n} = 0. 
\end{equation}
Thus, 
\begin{equation}
  \left\{
  \begin{array}{l}
  \mygr{A}{Q}_{n} = \mygr{A}{R}_{n} = 0 
  \\
  \mygr{M}{Q}_{n} = \mygr{M}{R}_{n} = 0 
  \end{array}
  \right.
  \quad\Rightarrow\quad
  \left\{
  \begin{array}{l}
  \mygr{A}{R}_{n+1} = 0
  \\
  \mygr{D}{QR}_{n+1} = 0 
  \end{array}
  \right.
  \quad\Rightarrow\quad
  \left\{
  \begin{array}{l}
  \mygr{A}{Q}_{n+1} = 0
  \\
  \mygr{A}{R}_{n+1} = 0 
  \end{array}
  \right.
\end{equation}
by virtue of (\ref{zero-AM-QR-a}) and (\ref{zero-AM-QR-b}). 
This proves the fact that if a pair $(Q_{n},R_{n})$ solves equations 
(\ref{hie-akns}) and the pair 
$(Q_{n+1},R_{n+1})$ is defined by $\mygr{M}{Q}_{n}=\mygr{M}{R}_{n}=0$, 
then it is another solution of (\ref{hie-akns}). 
In a similar way, one can prove that the pair 
$(Q_{n-1},R_{n-1})$ defined by $\mygr{M}{Q}_{n-1}=\mygr{M}{R}_{n-1}=0$ 
also solves (\ref{hie-akns}).

\section{Lattice representation of the DNLS hierarchy. \label{app-lr-dnls}} 

Along the lines of the previous appendix consider the quantities
\begin{eqnarray}
  \mytgr{M}{U}_{n}(\xi,\eta) & = & 
  \xi\mathbb{E}_{\xi} \, \mygr{M}{U}_{n}(\eta) 
  - 
  \eta\mathbb{E}_{\eta} \, \mygr{M}{U}_{n}(\xi) 
  \\
  \mytgr{M}{V}_{n}(\xi,\eta) & = & 
  \xi \mygr{M}{V}_{n}(\eta) 
  - 
  \eta \mygr{M}{V}_{n}(\xi) 
\end{eqnarray}
and
\begin{equation}
  \mathring\mygr{M}{UV}_{n}(\xi) 
  = 
  V_{n} \mygr{M}{U}_{n}(\xi) - \hat{U}_{n+1} \mygr{M}{V}_{n}(\xi) 
\end{equation}
which are related to $\mygr{A}{U,V}_{n}$ by 
\begin{equation}
  V_{n+1} \mygr{A}{U}_{n}(\xi,\eta) 
  + 
  \hat{\bar{U}}_{n} \mygr{A}{V}_{n+1}(\xi,\eta) 
  = 
  V_{n+1} \mytgr{M}{U}_{n}(\xi,\eta) 
  + 
  \hat{\bar{U}}_{n} \mytgr{M}{V}_{n}(\xi,\eta) 
\label{equ-AM-UV}
\end{equation}
and 
\begin{equation}
  \mygr{D}{UV}_{n}(\xi,\eta) - \mygr{D}{UV}_{n+1}(\xi,\eta) 
  = 
  \left( \mathbb{E}_{\xi} - 1 \right) \mathring\mygr{M}{UV}_{n}(\eta) 
  + 
  \left( 1 - \mathbb{E}_{\eta} \right) \mathring\mygr{M}{UV}_{n}(\xi) 
\label{equ-DM-UV}
\end{equation}
where $\mygr{D}{UV}_{n}$ is defined by (\ref{zero-D-UV}).
These identities can be used to prove the fact that if a pair $(U_{n},V_{n})$ 
solves equations (\ref{hie-dnls}) and the pairs 
$(U_{n \pm 1},V_{n \pm 1})$ are defined by 
$\mygr{M}{U}_{n}=\mygr{M}{V}_{n}=0$ and $\mygr{M}{U}_{n-1}=\mygr{M}{V}_{n-1}=0$, 
then they also solve (\ref{hie-dnls}).

\section{Generating relations for B\"acklund transformations. \label{app-grbt}}

Expanding the definition of the generating function $\mybacklund{G}_{n}(\xi)$, 
\begin{eqnarray}
  \xi^{-1} \mybacklund{G}_{n}(\xi)
  & = & 
  \hat{U}_{n} V_{n} - \hat{Q}_{n-1} R_{n} 
  - \xi \, \hat{Q}_{n-1} R_{n} \hat{U}_{n} V_{n} 
  \\ & = & 
  \hat{U}_{n} V_{n} - \hat{Q}_{n-1} R_{n} \mygamma{UV}_{n}(\xi) 
  \\ & = & 
  \hat{U}_{n} V_{n} \mygamma{QR}_{n-1}(\xi) - \hat{Q}_{n-1} R_{n} 
\end{eqnarray}
and taking the limit $\xi \to 0$, 
\begin{equation}
  \mybacklund{g}_{n} = 
  \lim_{\xi \to 0} \xi^{-1} \mybacklund{G}_{n}(\xi) = 
  U_{n} V_{n} - Q_{n-1}R_{n} 
\end{equation}
one can conclude than the functions
\begin{equation}
  \begin{array}{rcl} 
  \mygr{I}{Q}_{n} & = & 
  \hat{Q}_{n} / Q_{n}  - \hat{U}_{n+1} / U_{n+1} \mygamma{UV}_{n+1} 
  \\[2mm]
  \mygr{I}{R}_{n} & = & 
  R_{n} / \hat{R}_{n} -  V_{n} / \hat{V}_{n} \mygamma{UV}_{n} 
  \end{array}
\end{equation}
and 
\begin{equation}
  \begin{array}{rcl} 
    \mygr{I}{U}_{n} & = & 
    \hat{U}_{n} / U_{n} - \hat{Q}_{n-1} / Q_{n-1} \mygamma{QR}_{n-1} 
    \\[2mm]
    \mygr{I}{V}_{n} & = & 
    V_{n} / \hat{V}_{n} - R_{n} / \hat{R}_{n} \mygamma{QR}_{n-1}  
  \end{array}
\end{equation}
vanish together with $\mybacklund{G}_{n}(\xi)$:
\begin{equation}
  \mygr{I}{X}_{n} = 0 
  \;\mbox{mod}\; 
  \mybacklund{G}_{n}
  \qquad ({\scriptstyle X} = 
  {\scriptstyle Q}, {\scriptstyle R}, {\scriptstyle U}, {\scriptstyle V} ).
\end{equation}
The same is true, 
\begin{equation}
  \mytgr{I}{X}_{n} = 0 
  \;\mbox{mod}\; 
  \mybacklund{G}_{n} 
  \qquad ({\scriptstyle X} = 
  {\scriptstyle Q}, {\scriptstyle R}, {\scriptstyle U}, {\scriptstyle V} ),
\end{equation}
for the quantities
\begin{eqnarray}
  \mytgr{I}{Q}_{n} & = & 
  Q_{n} / \hat{Q}_{n} - \mygamma{UV}_{n+1} U_{n+1} / \hat{U}_{n+1}  
\label{backlund-tI-Q}
  \\[2mm]
  \mytgr{I}{R}_{n} & = & 
  \hat{R}_{n} / R_{n} - \mygamma{UV}_{n} \hat{V}_{n} / V_{n} 
\label{backlund-tI-R}
\end{eqnarray}
and
\begin{equation}
  \begin{array}{rcl} 
  \mytgr{I}{U}_{n} & = & 
  U_{n} / \hat{U}_{n} - \mygamma{QR}_{n-1} Q_{n-1} / \hat{Q}_{n-1}  
  \\[2mm]
  \mytgr{I}{V}_{n} & = & 
  \hat{V}_{n} / V_{n} - \mygamma{QR}_{n-1} \hat{R}_{n} / R_{n}. 
  \end{array}
\end{equation}
The quantities $\mygr{K}{QR}_{n}(\xi)$ and 
$\mygr{K}{UV}_{n}(\xi)$ defined by 
(\ref{backlund-K-QR}) and (\ref{backlund-K-UV})
can be presented as 
\begin{eqnarray}
  \mygr{K}{QR}_{n}(\xi) 
  & = & 
  \frac{ - U_{n+1}V_{n} \,\mygr{I}{U}_{n+1} 
         + \hat{U}_{n+1}\hat{V}_{n} \,\mygr{I}{V}_{n} } 
       { \hat{Q}_{n}R_{n} }
  - \frac{ \mybacklund{h}\mysup{UV}_{n} }{ Q_{n}R_{n} \mygamma{QR}_{n} } 
  + \frac{ \hat{\mybacklund{h}}\mysup{UV}_{n} }
         { \hat{Q}_{n}\hat{R}_{n} \mygamma{QR}_{n-1} } 
  \\[2mm] 
  \mygr{K}{UV}_{n}(\xi) 
  & = & 
  \frac{ - Q_{n}R_{n} \,\mygr{I}{Q}_{n} 
         + \hat{Q}_{n}\hat{R}_{n} \,\mygr{I}{R}_{n} } 
       { \hat{U}_{n+1}V_{n} }
  - \frac{ \mybacklund{h}\mysup{QR}_{n} }{ U_{n+1}V_{n} \mygamma{UV}_{n+1} } 
  + \frac{ \hat{\mybacklund{h}}\mysup{QR}_{n} }
         { \hat{U}_{n+1}\hat{V}_{n} \mygamma{UV}_{n} } 
\end{eqnarray}
where
\begin{eqnarray}
    \mygr{h}{QR}_{n} & = & 
    Q_{n}R_{n} - U_{n+1}V_{n} \left( 1 - U_{n}V_{n+1} \right)
    \\[2mm]
    \mygr{h}{UV}_{n} & = & 
    U_{n+1}V_{n} - Q_{n}R_{n} \left( 1 + Q_{n-1}R_{n+1}\right). 
\end{eqnarray}
It is easy to show that 
\begin{eqnarray}
    Q_{n-1}R_{n+1} \,\mygr{h}{QR}_{n} & = & 
    U_{n+1}V_{n} \,\mybacklund{h}_{n} - \mybacklund{g}^{h}_{n}
    \\[2mm]
    U_{n}V_{n+1} \,\mygr{h}{UV}_{n} & = & 
    - Q_{n}R_{n} \,\mybacklund{h}_{n} + \mybacklund{g}^{h}_{n} 
\end{eqnarray}
with
\begin{eqnarray}
    \mybacklund{g}^{h}_{n} & = & 
    U_{n}U_{n+1}V_{n}V_{n+1} - Q_{n-1}Q_{n}R_{n}R_{n+1}
    \\ & = &
    U_{n+1}V_{n+1} \,\mybacklund{g}_{n} 
    + Q_{n-1}R_{n} \,\mybacklund{g}_{n+1} 
    \\ & = &
    Q_{n}R_{n+1} \,\mybacklund{g}_{n} 
    + U_{n}V_{n} \,\mybacklund{g}_{n+1} 
\end{eqnarray}
which implies
\begin{equation}
  \mygr{h}{XY}_{n} = 0 
  \;\mbox{mod}\; 
  \mybacklund{g}_{n},\mybacklund{h}_{n}
  \qquad ({\scriptstyle XY} = {\scriptstyle QR}, {\scriptstyle UV} )
\end{equation}
and hence 
\begin{equation}
  \mygr{K}{XY}_{n}(\xi) = 0 
  \;\mbox{mod}\; 
  \mybacklund{G}_{n},\mybacklund{h}_{n}
  \qquad ({\scriptstyle XY} = {\scriptstyle QR}, {\scriptstyle UV}).
\end{equation}
In a similar way one can treat functions 
(\ref{backlund-L-Q})--(\ref{backlund-L-V}):
\begin{eqnarray}
  \mybacklund{L}\mysup{Q}_{n} 
  & = & 
  \frac{ 1 }{ U_{n}V_{n} } \left(
    \mybacklund{h}^{\scriptscriptstyle QR}_{n} 
    + \frac{ Q_{n} }{ Q_{n-1} } \,\mybacklund{g}_{n} 
  \right)
  \\
  \mybacklund{L}\mysup{R}_{n} 
  & = & 
  \frac{ 1 }{ U_{n+1}V_{n+1} } \left(
    \mybacklund{h}^{\scriptscriptstyle QR}_{n} 
    + \frac{ R_{n} }{ R_{n+1} } \,\mybacklund{g}_{n+1} 
  \right)
  \\
  \mybacklund{L}\mysup{U}_{n} 
  & = & 
  \frac{ 1 }{ Q_{n-1}R_{n} } \left(
    \mybacklund{h}^{\scriptscriptstyle UV}_{n} 
    - \frac{ U_{n+1} }{ U_{n} } \,\mybacklund{g}_{n} 
  \right)
  \\
  \mybacklund{L}\mysup{V}_{n} 
  & = & \displaystyle
  \frac{ 1 }{ Q_{n}R_{n+1} } \left(
    \mybacklund{h}^{\scriptscriptstyle UV}_{n} 
    - \frac{ V_{n} }{ V_{n+1} } \,\mybacklund{g}_{n+1} 
  \right).
\end{eqnarray}
Hence, 
\begin{equation}
  \mygr{L}{X}_{n} = 0 
  \;\mbox{mod}\; 
  \mybacklund{G}_{n},\mybacklund{h}_{n}
  \qquad ({\scriptstyle X} = 
  {\scriptstyle Q}, {\scriptstyle R}, {\scriptstyle U}, {\scriptstyle V} ). 
\end{equation}
Finally, the properties of the functions $\mygr{F}{QU}_{n}$ and 
$\mygr{F}{RV}_{n}$ that are used in the body of the paper 
(in particular, equations (\ref{equ-A-QU})--(\ref{equ-M-RV})) follow from the identities 
\begin{eqnarray}
    \mygr{F}{QU}_{n}(\xi)  
    & = & 
    - \frac{ \hat{Q}_{n} }{ \mygamma{QR}_{n} } \,\mygr{M}{R}_{n} 
    - Q_{n} R_{n+1} \,\mygr{I}{U}_{n+1} 
    - \frac{ \hat{U}_{n+1} }{ U_{n+1} } \mybacklund{g}_{n+1} 
    + \hat{\mybacklund{g}}_{n+1}      
\label{prop-F-a}
    \\[2mm] 
    & = & 
    - \frac{ \hat{U}_{n+1} }{ \mygamma{UV}_{n} } \,\mygr{M}{V}_{n} 
    + \frac{ \xi V_{n} }{ \hat{V}_{n} \mygamma{UV}_{n} } 
      \,\hat{\mybacklund{h}}\mysup{QR}_{n} 
    + \xi \hat{Q}_{n} \hat{R}_{n} \,\mygr{I}{R}_{n} 
  \\[4mm] 
    \mygr{F}{RV}_{n}(\xi)  
    & = & 
    - \frac{ R_{n} }{ \mygamma{QR}_{n-1} } \,\mygr{M}{Q}_{n-1} 
    - \hat{Q}_{n-1} \hat{R}_{n} \,\mygr{I}{V}_{n} 
    + \mybacklund{g}_{n}      
    - \frac{ V_{n} }{ \hat{V}_{n} } \,\hat{\mybacklund{g}}_{n} 
    \\[2mm] 
    & = & 
    - \frac{ V_{n} }{ \mygamma{UV}_{n+1} } \,\mygr{M}{U}_{n} 
    + \frac{ \xi \hat{U}_{n+1} }{ U_{n+1} \mygamma{UV}_{n+1} } 
      \,\mybacklund{h}\mysup{QR}_{n} 
    + \xi Q_{n} R_{n} \,\mygr{I}{Q}_{n} 
\label{prop-F-b}
\end{eqnarray} 
and 
\begin{eqnarray}
  \frac{ \mygr{A}{U}_{n+1} }{ \hat{\bar{U}}_{n+1} } 
  - 
  \frac{ \mygr{A}{Q}_{n} }{ \hat{\bar{Q}}_{n} } 
  & = & 
  \xi\eta\left[ \mygr{F}{QU}_{n}(\xi) - \mygr{F}{QU}_{n}(\eta) \right] 
  + 
  \xi\mathbb{E}_{\xi} \mytgr{I}{Q}_{n}(\eta) 
  - 
  \eta\mathbb{E}_{\eta} \mytgr{I}{Q}_{n}(\xi) 
\label{prop-F-c}
  \\ 
  \frac{ \mygr{A}{V}_{n} }{ V_{n} } 
  - 
  \frac{ \mygr{A}{R}_{n} }{ R_{n} } 
  & = & 
  \xi\eta\left[ 
    \mathbb{E}_{\eta} \mygr{F}{RV}_{n}(\xi) 
  - \mathbb{E}_{\xi}  \mygr{F}{RV}_{n}(\eta) 
  \right] 
  + \xi  \mytgr{I}{R}_{n}(\eta) 
  - \eta \mytgr{I}{R}_{n}(\xi). 
\label{prop-F-d}
\end{eqnarray}
that can be verified directly.

\section*{References.}


\begin{thebibliography}{99}

\bibitem{N85}
  Newell A.C.,
  1985,
  \textit{Solitons in mathematics and physics}
  (Philadelphia: Society for Industrial and Applied Mathematics).


\bibitem{WKI1979}
  Wadati M., Konno K. and Ichikawa Y.-H., 
  1979,
  A generalization of inverse scattering method.
  \textit{J. Phys. Soc. Japan}, \textbf{46}, 1965--1966. 

\bibitem{KSK80}
  Kawata T., Sakai J.-I. and Kobayashi N.,
  1980,
  Inverse method for the mixed nonlinear Schr\"odinger equation and soliton solutions.
  \textit{J. Phys. Soc. Japan}, \textbf{48}, 1371--1379.

\bibitem{IKWS80}
  Ichikawa Y.H., Konno K., Wadati M. and Sanuki H.,
  1980,
  Spiky soliton in circular polarized Alfv\'en wave.
\textit{J. Phys. Soc. Japan}, \textbf{48}, 279--286.

\bibitem{FNR83}
  Flaschka H., Newell A.C. and Ratiu T.,
  1983,
  Kac-moody lie algebras and soliton equations: II. Lax equations associated with $A_{1}^{(1)}$. 
  \textit{Physica D}, \textbf{9}, 300--323. 

\bibitem{DM06a}
  Dimakis A. and M\"uller-Hoissen F., 
  2006,
  From AKNS to derivative NLS hierarchies via deformations of associative products.
  \JPA, \textbf{39}, 14015--14033.

\bibitem{I82}
  Ishimori Y., 
  1982,
  A relationship between the Ablowitz-Kaup-Newell-Segur and Wadati-Konno-Ichikawa 
  schemes of the inverse scattering method.
\textit{J. Phys. Soc. Japan}, \textbf{51}, 3036--3041. 

\bibitem{WS83}
  Wadati M. and Sogo K., 
  1983, 
  Gauge transformations in soliton theory. 
  \textit{J. Phys. Soc. Japan}, \textbf{52}, 394--398.

\bibitem{HW94}
  Hisakado M. and Wadati M., 
  1994, 
  Gauge transformations among generalised nonlinear Schr\"o\-dinger equations. 
  \textit{J. Phys. Soc. Japan}, \textbf{63}, 3962--3966.

\bibitem{V02}
  Vekslerchik V.E., 
  2002,
  Functional representation of the Ablowitz-Ladik hierarchy. II.
  \textit{J. Nonlin. Math. Phys.}, \textbf{9}, 157--180.

\bibitem{DM06b}
  Dimakis A., M\"uller-Hoissen F., 
  2006, 
  Functional representations of integrable hierarchies. 
  \JPA,  \textbf{39}, 9169--9186.

\bibitem{PV02}
  Pritula G.M. and Vekslerchik V.E., 
  2002,
  Conservation laws for the nonlinear Schr\"odinger equation in Miwa variables. 
  \textit{Inverse Problems}, \textbf{18}, 1355--1360.

\bibitem{AKNS74}
  Ablowitz M.J., Kaup D.J., Newell A.C. and Segur H., 
  1974, 
  The inverse scattering transform -– Fourier analysis for nonlinear problems, 
  \textit{Stud. Appl. Math.}, \textbf{53}, 249--315.

\bibitem{ZS71}
  Zakharov V.E. and Shabat A.B., 
  1971, 
  Exact theory of two-dimensional self-focusing and one-dimensional self-modulation of waves in nonlinear media, 
  \textit{ZhETF}, \textbf{61}, 118--134 
  (English translation in: \textit{Sov. Phys. JETP}, 1972, \textbf{34}, 62-–69). 

\bibitem{ZS73}
  Zakharov V.E. and Shabat A.B., 
  1973, 
  Interaction between solitons in a stable medium, 
  \textit{ZhETF}, \textbf{64}, 1627--1639 
  (English translation in: \textit{Sov. Phys. JETP}, \textbf{37}, 823--828).

\bibitem{KN78}
  Kaup D.J. and Newell A.C., 
  1978, 
  An exact solution for a derivative nonlinear Schr\"odinger equation. 
  \textit{J. Math. Phys.}, \textbf{19}, 798-801. 

\bibitem{CLL79}
  Chen H.H., Lee Y.C. and Liu C.S., 
  1979, 
  Integrability of nonlinear Hamiltonian systems by inverse scattering method. 
  \textit{Physica Scripta}, \textbf{20}, 490--492.

\bibitem{GI83}
  Gerdjikov V.S., Ivanov M.I.,
  1983,
  The quadratic bundle of general form and the nonlinear evolution equations. II. Hierarchies of Hamiltonian structures.    \textit{Bulgarian J. Phys.}, \textbf{10}, 130--143  (in Russian).

\bibitem{V11}
  Vekslerchik V.E.,
  2011,
  Lattice representation and dark solitons of the Fokas-Lenells equation. 
  \textit{Nonlinearity}, \textbf{24}, 1165--1175.

\end{thebibliography}
\end{document}